%% file: secrecy_interference.tex
\documentclass[10pt,journal]{IEEEtran}
\usepackage{cite}
\usepackage{psfrag}
\usepackage[pdf]{graphicx}
\usepackage[latin1]{inputenc}
\usepackage[T1]{fontenc}
\usepackage{amsmath,amsfonts,amsbsy,amssymb}
\usepackage{mathabx}
\usepackage{mathrsfs}
\usepackage[nolist]{acronym}
\usepackage{tabularx}

\usepackage{amssymb}
\usepackage{amsmath}
\usepackage{graphicx}
\usepackage[latin1]{inputenc}
\usepackage{graphicx}
\usepackage{cite}
\usepackage{multirow}
\usepackage{wasysym}
\usepackage{multirow}
\usepackage{float}

\usepackage{color}

\usepackage[absolute,overlay]{textpos}
\usepackage{tikz}
\usetikzlibrary{topaths}
\usetikzlibrary{shapes,trees,decorations,decorations.pathreplacing,decorations.pathmorphing}
\usetikzlibrary{arrows}
\usetikzlibrary[shadows]
\usetikzlibrary{positioning}
\usetikzlibrary{matrix}
\definecolor{darkblue}{rgb}{.1,.1,.6}

\usepackage[psfixbb,tightpage,displaymath,floats]{preview}
\PreviewEnvironment{tikzpicture}
\usepackage[colorlinks,bookmarksopen,bookmarksnumbered,citecolor=red,urlcolor=red]{hyperref}

\newtheorem{definition}{Definition}

\newtheorem{theorem}{Theorem}

\newcommand{\review}[1]{#1}

\begin{document}

\title{\review{On the Secrecy of Interference-Limited Networks under Composite Fading Channels}}
\author{
	\IEEEauthorblockN{Hirley Alves~\IEEEmembership{Student Member, IEEE}, Carlos H.~M. de~Lima~\IEEEmembership{Member, IEEE}, Pedro. H. J. Nardelli~\IEEEmembership{Member, IEEE}, Richard Demo Souza~\IEEEmembership{Senior Member, IEEE} and Matti Latva-aho~\IEEEmembership{Senior Member, IEEE}} 
	\thanks{H. Alves, C. H. M. de Lima, P. H. J. Nardelli and M. Latva-aho are with the Centre for Wireless Communications (CWC), University of Oulu, Finland. Contact: halves@ee.oulu.fi. 
	
	R. D. Souza and H. Alves are with Federal University of Technology - Paraná (UTFPR), Curitiba, Brazil. 
	
	C. H. M. de Lima is also with São Paulo State University (UNESP), São João da Boa Vista, Brazil. 
	
	The authors also would like to thank Aka and Infotech Oulu Graduate School from Finland, and CNPq and Special Visiting Researcher fellowship CAPES 076/2012 from Brazil.}
}
%

\maketitle
%

\input{acronym}

\begin{abstract}
	This paper deals with the secrecy capacity of the radio channel in interference-limited regime.
	We assume that interferers are uniformly scattered over the network area according to a Point Poisson Process and the channel model consists of path-loss, log-normal shadowing and Nakagami-m fading. 
	Both the probability of non-zero secrecy capacity and the secrecy outage probability are then derived in closed-form expressions using tools of stochastic geometry and higher-order statistics.
	Our numerical results show how the secrecy metrics are affected by the disposition of the desired receiver, the eavesdropper and the legitimate transmitter.
	\vspace{-2mm}
\end{abstract}

\begin{keywords}
Composite channel, secrecy capacity, secrecy outage probability,  stochastic geometry 
\vspace{-2mm}
\end{keywords}

\acresetall
%

\section{Introduction}
\label{SSE:INTRODUCTION}
Due to their broadcast nature, wireless communications are susceptible to security issues since non-intended nodes within the communication range of a given transmitter can overhear the transmission and possibly extract private information \cite{ShiuPHYTutoralWC2011}.
To ensure confidentiality, cryptographic techniques (usually implemented in higher layers) depend on secret keys and also rely on the limited computational power of eavesdroppers and on the reliability guaranteed by channel coding at the \ac{PHY}. 
However, future wireless systems tend to be deployed in large scale with ubiquitous coverage, dynamic operation and computational powerful devices, making encrypted communication through secret keys expensive and difficult to achieve. 
\acs{PHY} security has reemerged \cite{BlochTIT2008,MukherjeeCSurveyTut2014,ShiuPHYTutoralWC2011,BassilySPM2013} as a viable alternative to enhance the robustness and reduce the complexity of conventional cryptography systems since \acs{PHY} offers unbreakable and quantifiable secrecy in confidential bps/Hz, regardless of the eavesdropper's computational power.

\acs{PHY} security dates back to 1975, when Wyner in his pioneering work \cite{ART:WYNER-BELL75} introduced the wire-tap channel which is composed of a pair of legitimate users, known as Alice (transmitter) and Bob (receiver), communicating in the presence of an eavesdropper known as Eve.
Alice and Bob communicate through the main channel in the presence of Eve, who perceives a degraded version of the message sent to Bob through the eavesdropper channel.
In this context it is proved that there exist codes which guarantee both low error probabilities and a certain degree of confidentiality.

Later in \cite{ART:LEUNG-IT78}, it was demonstrated that the secrecy capacity of the Gaussian wire-tap channel can be defined as the difference between the capacity of the main channel and the eavesdropper channel considering that the eavesdropper channel is noisier than the main channel.
The wire-tap channel is extended to a fading scenario and secrecy outage probability is then characterized in \cite{BlochTIT2008,MukherjeeCSurveyTut2014}. 
Cooperative and jamming techniques are reviewed in \cite{BassilySPM2013} and diversity schemes are assessed in \cite{ART:HIRLEY_SPL12}.

Such recent works, however, focus on a small number of nodes, whose results provide few insights how \acs{PHY} security performs in large-scale  networks \cite[Sec. VIII-C]{MukherjeeCSurveyTut2014}.
Different from the point-to-point communication wherein secrecy is guaranteed by keys or tokens, large-scale deployments impose new challenges to ensure secure communication, due to the complexity of distributing and maintaining secret keys \cite{edsilva}.

Security in large-scale networks are also affected by the spatial distribution of the interferer nodes and eavesdroppers (for more details, refer to \cite[Sec. VIII-C]{MukherjeeCSurveyTut2014} and the references therein).
Early contributions aim at characterizing the scaling laws of secrecy capacity \cite{ART:GAMAL-IT12} and the network connectivity \cite{ART:BARROS1-IFS12} of randomly scattered nodes, or at computing the secure area spectral efficiency for predefined quality requirements.
Recent papers have also shown that PHY-security can compensate for the vulnerability of wireless communications and reduce its implementation complexity by exploiting the spatial-temporal characteristics of the wireless medium \cite{ART:ZHOU-WCOM11}. 

In contrast, we characterize here the secrecy capacity of large-scale networks in the presence of uncoordinated interference by applying a general framework that jointly considers nodes spatial distribution and composite-fading channel, a model introduced in \cite{ART:LIMA-TWC12A,Lima2013}.
Besides, as most of existing works consider thermal noise alone to compute the secrecy capacity, this paper also contributes to the literature by providing a framework that allows for evaluating the aggregate interference and hence computing the non-zero secrecy capacity and the secrecy outage probability in closed-form.

Built upon the analytic results from \cite{ART:LIMA-TWC12A,Lima2013}, we  assess the joint effects of transmitter-receiver and transmitter-eavesdropper relative distances and density interfering nodes on such capacity metrics.
Considering a scenario where interferers follow a Point Poisson Process and the channel model consists in path-loss, log-normal shadowing and Nakagami-m fading,  we show under which circumstances non-zero secrecy capacity is possible and how the secrecy outage probability behaves in terms of such dynamics. 

%
%
\section{System Model}\label{SSE:SYSTEM_MODEL}
\vspace{3mm}

This section describes the system model used here, following the basic concepts introduced in \cite[Sec.II]{ART:LIMA-TWC12A}.
\review{To begin with, the tagged legitimate pair is defined as the reference link (transmitter-receiver) so as to compute the aggregate CCI and performance metrics for the scenario under consideration.}
Specifically the mutual information of the tagged legitimate pair and  its related eavesdropper are determined
based on their instantaneous \ac{SIR} values.
We adopt the notion for secrecy capacity, probability of existence of non-zero secrecy capacity and secrecy outage probability as in \cite[Sec. II]{MukherjeeCSurveyTut2014}. 

We consider large-scale wireless networks where legitimate transmitters (interferer nodes) constitute a homogeneous \acp{PPP} $\Phi$ with density
$\lambda$ [TXs/m$^2$] in $\mathbb{R}^2$ \cite{ART:LIMA-TWC12A,Lima2013}.
%
%
Transmitters communicate using antennas with omni-directional radiation pattern and
fixed transmit power.

Let us now consider the set of legitimate transmitters in an arbitrary region
$\mathcal{R}$ of area $A$ which thus follows a Poisson distribution with parameter
$\lambda A$.
We then assume the (composite) fading effect as a random mark associated with each
point of $\Phi$.
Using \textup{Marking theorem}, the resulting process $	{\widetilde{\Phi}} = \left\{ \left( \varphi, x \right); \varphi \in \Phi \nonumber
	\right\}$ corresponds to a \ac{MPP} on the product space $\mathbb{R}^2 \times
\mathbb{R}^{+}$, whose points (transmitter locations) $\varphi$ belong to the process $\Phi$ and the random variable $x$ refers to the corresponding squared-envelop of the composite fading, as presented later.

\review{
The scenarios under study are interference-limited and hence the thermal noise is negligible in comparison to the resulting \ac{CCI} \cite{ART:LIMA-TWC12A,Lima2013}. We recall that as pointed out by \cite{ART:GHASEMI-JSAC08} the aggregate interference dominates over AWGN noise and the distribution of the aggregate interference is positively skewed and heavy tailed, which suggests a Log-Normal distribution \cite{ART:GHASEMI-JSAC08}.}
\review{We assume the high mobility random walk model, so each observation period can be analyzed as an independent realization of the \ac{MPP}\cite{BaccelliNOW2010}.} 
\begin{figure}[!t]
	\centering
	\includegraphics[width=\columnwidth]{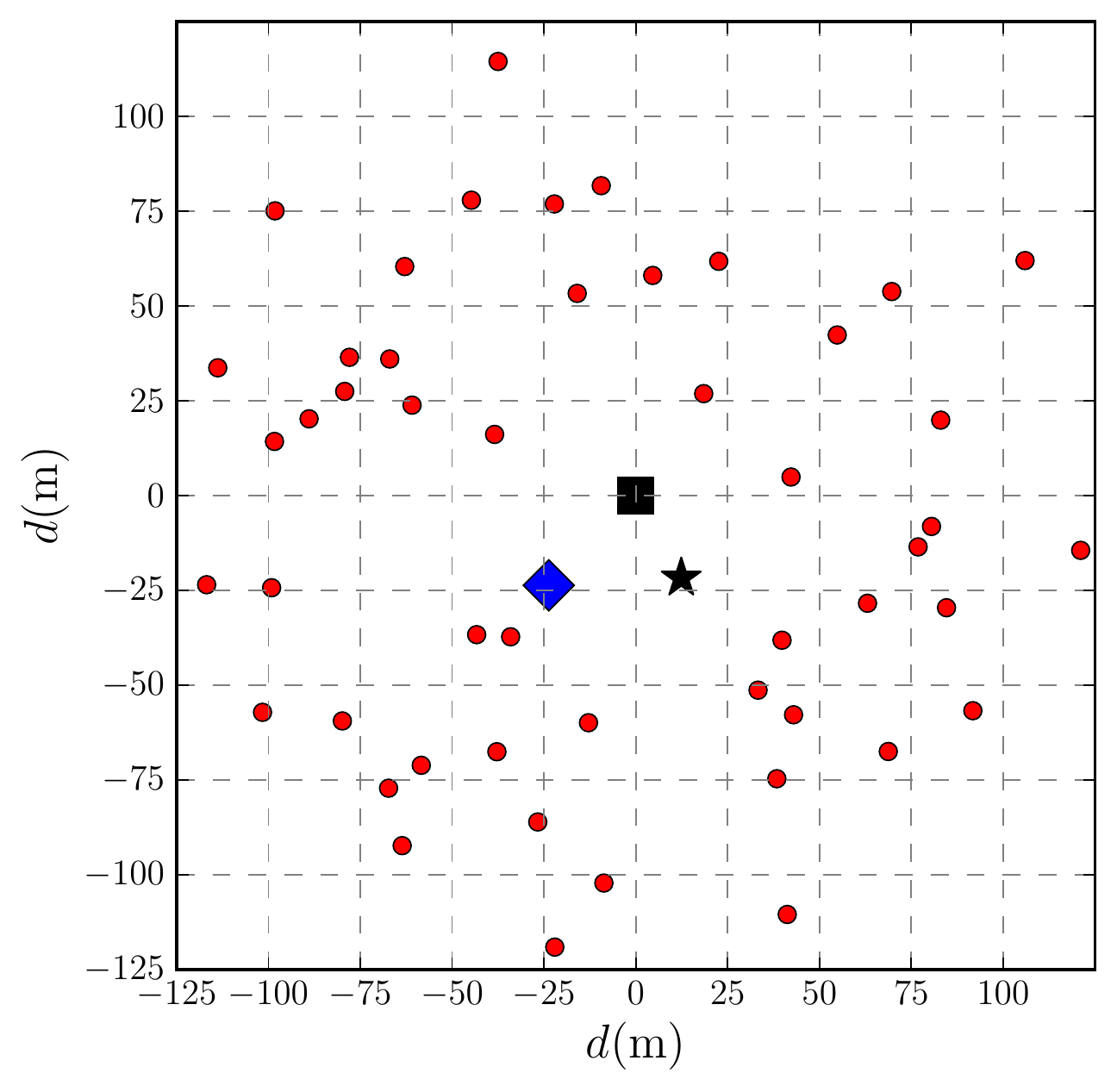}	
	%
	\caption{\review{Example of the network model. Legitimate pair is represented in black (square and star) and are separated by a distance $d_l$, while Eve is depicted in blue (diamond) and at distance $d_e$ from the tagged transmitter. Note that the interferers are in red (circles) and that the aggregate power of the interferers disrupt the communication of the tagged receiver and the eavesdropper. }}
	\label{FIG:NET_MODEL}
\end{figure}

Fig.~\ref{FIG:NET_MODEL} depicts the network model in a 2D grid. 
\review{Notice that the legitimate pair -- the tagged transmitter and receiver -- is represented in black (star and square, respectively) and are separated by a distance of $d_l$.}
The field of interferers is denoted by red circles while the eavesdropper is represented by the blue diamond. 
\review{ We assume that the eavesdropper is at a known distance $d_e$ from the tagged transmitter.}
Through the framework introduced in \cite{ART:LIMA-TWC12A,Lima2013}, we are able to compute the aggregate interference seen by the receivers. 

Radio links are subject to distance-dependent path-loss and shadowed fading, which is assumed to
be independent over distinct network entities and positions.
\review{
A signal strength decay function describes the average power attenuation (unbounded model) as $l\negthinspace\left( d_i \right) = d_i^{-\alpha}$, where $\alpha$ is the path loss exponent and $d_i$ represents the distance between a transmitter-receiver pair with $i \in \{l \,,\,e \}$, which can be either a legitimate receiver or the eavesdropper.}

\review{
Each interferer then disrupts the communication of the tagged receiver with a component given by $p\,\thinspace l\negthinspace\left( d \right) x$, where $p$ represents the interferer's transmit power, $d$ is the separation distance from an interferer to the tagged receiver or eavesdropper. 
From this assumption, we can compute an approximation to the distribution of the aggregate interference caused by all active transmitters, defined by the MPP, as presented in the next section.
}

\review{
The received squared-envelop due to multipath fading and shadowing is represented
by a \ac{RV} $X \in \mathbb{R}^+$ with \ac{CDF}, $F_X\negthinspace\left( x \right)$,  and \ac{PDF}, $f_X\negthinspace\left( x \right)$. Then, the composite distribution of the received squared-envelop due to \ac{LN} shadowing and Nakagami-$m$ fading has a Gamma-\ac{LN} distribution,  whose \ac{PDF} is given as \cite{ART:LIMA-TWC12A,Lima2013}:
}

%
\begin{align}\label{EQ:GAMA_LOGNORMAL_PDF}
	f_X(x) &= \int\limits_{0}^{\infty}\left(\frac{m}
	{\omega}\right)^m\frac{x^{m-1}} {\Gamma(m)}
	\exp\left(-\frac{m}{\omega}x\right) \\
	&\times \frac{\xi}{\sqrt{2 \pi} \sigma
	\omega}\exp{\left[ -\frac{\left ( \xi\ln{\omega} -\mu_{\Omega_p}
	\right)^2}{2 \sigma_{\Omega_p}^2} \right]}\mathrm{d}\omega \nonumber \, . 
\end{align} 

{\noindent In this case, 
$m$ is the shape parameter of the Gamma (Nakagami-$m$) distribution, $\Gamma(\cdot)$ is the gamma function
\cite[Eq.~8.310-1]{BOOK:GRADSHTEYN-ACADEMIC07}, $\xi = \ln \left( 10
\right)/10$, $\Omega_p$ is the mean squared-envelop, $\mu_{\Omega_p}$  and
$\sigma_{\Omega_p}$ is the mean and standard deviation of $\Omega_p$,
respectively.} 

\review{Moreover, Ho and St\"{u}ber show in \cite{ART:HO-ACM95} that a composite Gamma--\ac{LN}
distribution can be approximated by a single \ac{LN} distribution with mean and
variance (in logarithmic scale) given by $\mu_{\mathrm{dB}} = \xi \left [
\psi\left ( m \right ) - \ln\left ( m \right)\right ] + \mu_{\Omega_p}$ and
$\sigma_{\mathrm{dB}}^2 = \xi^{2}\zeta\left ( 2, m \right) +
\sigma_{\Omega_p}^2$, where $\psi\left ( m \right )$ is the Euler psi function
\cite[Eq.~8.360-1]{BOOK:GRADSHTEYN-ACADEMIC07} and $\zeta\left ( 2, m \right )$
is the generalized Riemann zeta function
\cite[Eq.~9.551]{BOOK:GRADSHTEYN-ACADEMIC07}.}


\section{Secrecy Capacity Analysis}\label{SSE:L1_SECURITY}
The cumulant-based framework from \cite[Secs. II, IV and VII]{ART:LIMA-TWC12A} and \cite[Sec.VI]{Lima2013} is now used to asses achievable levels of secrecy and the resulting performance of legitimate link.

\review{Suppose that the legitimate transmitter has only CSI of the desired receiver, which is known as passive eavesdropping \cite{ShiuPHYTutoralWC2011, ART:HIRLEY_SPL12}. In such case, we resort to a probabilistic view of security in order to characterize the probability of information leakage to the eavesdropper.}
\review{
Then, in order to protect the transmission from an inimical attack, we consider the use of a wiretap code with  $2^{n{R}}$ codewords, where ${R}$ is made equal to the instantaneous capacity of the legitimate channel, namely $C_l$ \cite{BlochTIT2008}. Then, the number of codewords per bin is set equal to $2^{n{R}_e}$, where ${R}_e$ represents the eavesdropper's equivocation rate. Thus, a fixed secure transmission rate is attained as ${R}_{s}= {R} - {R}_e = C_l - {R}_e$, which implies that ${R}_e= C_l - R_s$ varies according to the legitimate channel condition. Therefore, as introduced in \cite{BlochTIT2008}, an outage event occurs when ${R}_s$ exceeds the difference between the instantaneous capacities of the legitimate and the eavesdropper channels, thus $\Pr\left[ [C_l - C_e ]^+ < {R}_s\right]$, with $[\cdot]^+ \triangleq \max\{\cdot,0\}$.}

\review{
Let us first characterize the probability of existence of non-zero secrecy capacity ($\Pr \left[ C_l > C_e \right]$) when the legitimate link experiences interference from concurrent transmissions. We use $\Gamma_l$ and $\Gamma_e$ to denote the SIR of the legitimate and eavesdropper links, respectively.
%
Then, }the following expression shows the secrecy capacity of shadowed fading channels
\begin{align}
	C_{s} &= \left[ C_{l} - C_{e}\right]^{+}\hspace{-0em}
	= \left[\log {\left ( 1 + \Gamma_l \right)} - \log{\left ( 1 + \Gamma_e \right)}\right]^{+} \hspace{-0em},
	\label{EQ:SECRECY_CAPACITY}
\end{align}
%
%
{\noindent The distributions of $\Gamma_l$ and $\Gamma_e$ can be recovered using \cite[Th.1]{Lima2013}.}
%

\begin{theorem}
	For the system model described in this paper and the
	distances $d_l$ (from the receiver)  and $d_e$ (from the eavesdropper) to the
	tagged legitimate transmitter, the probability of existence of non-zero secrecy capacity is 
	\begin{align}
		\Pr \left[ C_s > 0 \right] = \operatorname{Q}\left[\frac{\mu_e
		-\mu_l}{\sqrt{\sigma_l^2 + \sigma_e^2}}\right]
		\label{EQ:SECRECY_CAPACITY_EXISTENCE}
	\end{align}
	\label{THEOREM:SECRECY_CAPACITY_EXISTENCE}
	\vspace{-1mm}
\end{theorem}
\begin{IEEEproof}
		The non-zero secrecy capacity probability is
		\begin{align}
			\Pr \left[ C_s > 0 \right] =& \Pr \left[ \Gamma_l > \Gamma_e \right] 
			= \hspace{-1ex} \int\limits_{0}^{\infty}
			\int\limits_{0}^{\gamma_l} \hspace{-1mm} f_l \left ( \gamma_l \right ) f_e \left (
			\gamma_e \right) \mathrm{d}\gamma_e\, \mathrm{d}\gamma_l ,
			\label{EQ:B_S1}
		\end{align}
		{\noindent where $C_s$ is defined in \eqref{EQ:SECRECY_CAPACITY}, the \ac{SIR} \acp{PDF} $f_l \left ( \gamma_l \right )$ and $f_e \left ( \gamma_e
				\right)$ of the legitimate and
		eavesdropper nodes follow $\mathsf{Lognormal}\left( \mu_l, \sigma_l^2
		\right)$ and $\mathsf{Lognormal}\left( \mu_e, \sigma_e^2 \right)$.}
		Integrating in terms of $\gamma_e$, we have
		\begin{align}
			\Pr \left[ \Gamma_l > \Gamma_e \right] =& \int\limits_{0}^{\infty}
			\frac{1}{2} \operatorname{Erfc} \left[ \frac{\mu_e - \log \left(
			\gamma_e \right)}{\sqrt{2} \sigma_e} \right] f_l \left ( \gamma_l
			\right ) \mathrm{d}\gamma_l.
			\label{EQ:B_S2}
		\end{align}
		
		Thereafter, we substitute $\eta = {\left( \mu_e - \log \gamma_e \right)} /
		{\sqrt{2} \sigma_e}$ in \eqref{EQ:B_S2} and adjust the limits of
		integration accordingly to obtain
		\begin{align}
			\Pr \left[ \Gamma_l > \Gamma_e \right] =&
			\int\limits_{-\infty}^{\infty} \frac{e^{-\eta^2}}{2 \sqrt{\pi}}
			\operatorname{Erf}\left[ \frac{-\mu_l + \mu_e + \sqrt{2} \alpha
			\sigma_l}{\sqrt{2} \sigma_e} \right] \mathrm{d}\eta.
			\label{EQ:A1_S3}
		\end{align}
		
		Eq. \eqref{EQ:SECRECY_CAPACITY_EXISTENCE} is obtained from
		\eqref{EQ:A1_S3} by using \cite[Eq. 8.259-1]{BOOK:GRADSHTEYN-ACADEMIC07}.
\end{IEEEproof}

We now need to identify the circumstances whereby secrecy is compromised by defining the secrecy outage probability.
\begin{definition}{\textbf{Secrecy Outage Probability}}
	is defined as the probability that the instantaneous secrecy capacity $C_s$
	does not match the target secrecy rate $R_s > 0$ and is expressed as \cite{MukherjeeCSurveyTut2014}:
	\begin{align}
		&\Pr \left[ C_s < R_s \right] = \Pr \left[ C_s < R_s \,|\, \Gamma_l >
		\Gamma_e \right] \Pr \left[ \Gamma_l > \Gamma_e \right] + \nonumber \\
		&\qquad \qquad \qquad \,\,\quad \Pr \left[
		C_s < R_s \,|\, \Gamma_l \leq \Gamma_e \right] \Pr \left[ \Gamma_l \leq
		\Gamma_e \right].
		\label{EQ:OUTAGE_SECRECY_CAPACITY_1}
	\end{align}
\end{definition}

%
\begin{theorem}
	For the system model described in this paper, the secrecy outage probability with respect to the legitimate link and
	an arbitrary eavesdropper is given by
	\begin{align}\label{EQ:SECRECY_OUTAGE_PROBABILITY}
		&\Pr \left[ C_s < R_s \right] = \frac{1}{2} -
		\sum_{n=1}^{N} \frac{\omega_n}{2 \sqrt{\pi}} \\
		&\times 
		\operatorname{Erf}\left[\frac{\mu_l - \log\left[-1 + 2^{R_s} + 2^{R_s}
		\exp{\left( \mu_e - \sqrt{2} \eta_n \sigma_e \right)}\right]}{\sqrt{2}
		\sigma_l}\right]. \nonumber
		%
		%
	\end{align}
	\label{THEOREM:OUTAGE_SECRECY_CAPACITY}
\end{theorem}

\begin{IEEEproof}
	Let us start evaluating each summand in
		\eqref{EQ:OUTAGE_SECRECY_CAPACITY_1} separately.
		Recall from Theorem \ref{THEOREM:SECRECY_CAPACITY_EXISTENCE} that
		$\Pr \left[ \Gamma_l > \Gamma_e  \right] = \operatorname{Q} \left[ \left(
		\mu_e - \mu_l \right) / \sqrt{\sigma_l^2 + \sigma_e^2} \right]$	and since
		$R_s > 0$, it follows that $ \Pr \left[ C_s < R_s \,|\, \Gamma_l
		\leq \Gamma_e \right] = 1$.
		
		Then, we  rewrite the secrecy capacity
		in terms of its \ac{SIR} distribution and then proceed as
		\begin{align}
			&\Pr \left[ C_s < R_s \,|\, \Gamma_l > \Gamma_e \right]= \nonumber \\
			\quad &= \Pr \left[
			\log_2 \left( 1 + \Gamma_l \right) - \log_2 \left( 1 + \Gamma_e \right) <
			R_s \,|\, \Gamma_l > \Gamma_e \right] \nonumber \\
			&= \Pr \left[ \Gamma_l < 2^{R_s} \left( 1 + \Gamma_e \right) - 1 \,|\,
			\Gamma_l > \Gamma_e \right].
			\label{EQ:C_S1}
		\end{align}
		
		Under the assumption that legitimate and eavesdropper channels are
		independent, one computes \eqref{EQ:C_S1} as follows.
		\begin{align}
			&\Pr \left[ C_s < R_s \,|\, \Gamma_l > \Gamma_e \right] = \nonumber \\
			\quad &= \frac{1}{\Pr
			\left[ \Gamma_l > \Gamma_e \right]} \int\limits_{0}^{\infty}
			\int\limits_{0}^{ \epsilon \left( 1 + \gamma_e \right) - 1} f_l \left (
			\gamma_l \right ) f_e \left ( \gamma_e \right) \mathrm{d}\gamma_l\,
			\mathrm{d}\gamma_e,
			\label{EQ:C_S2}
		\end{align}
		{\noindent where $\epsilon = 2^{R_s}$.}
		After integrating over $\gamma_l$, we attain
		\begin{align}
			&\Pr \left[ C_s < R_s \,|\, \Gamma_l > \Gamma_e \right] = 
			\frac{1}{\Pr 	\left[ \Gamma_l > \Gamma_e \right]} (\Xi_1 - \Xi_2).
			\label{EQ:C_S3}
		\end{align}
		%
		
		Employing the same method used in the previous proof\cite[Eq. 8.259-1]{BOOK:GRADSHTEYN-ACADEMIC07}, we have
		\begin{align}
			\Xi_1 \hspace{-1mm}  &= \hspace{-1ex} \int\limits_{0}^{\infty} \hspace{-1mm}  f_e \left ( \gamma_e
			\right) \operatorname{Erf} \left[ \frac{\mu_l -
			\log{\gamma_e}}{\sqrt{2} \sigma_l} \right] \mathrm{d}\gamma_e \hspace{-1mm} 
			= \hspace{-1mm} 
			\operatorname{Q} \left[ \frac{\mu_l - \mu_e}{\sqrt{\sigma_l^2 -
			\sigma_e^2}} \right],
			\label{EQ:C_S4}
		\end{align}
		where the second term is integrated using Gauss-Hermite quadrature
		\cite{BOOK:ABRAMOWITZ-DOVER03} and $\eta = \left(
		\mu_e - \log{\gamma_e} \right)/\sqrt{2}\sigma_e$, so that
		%
	%
		\begin{align}
			\Xi_2 &=  \int\limits_{0}^{\infty} f_e \left ( \gamma_e
			\right) \operatorname{Erf} \left[ \frac{\mu_l - \log \left( -1 + \epsilon +
			\epsilon \gamma_e \right)}{\sqrt{2} \sigma_l}\right] \mathrm{d}\gamma_e
			\nonumber \\
			%
			%
			&= \sum_{n=1}^{N}\frac{\omega_n}{2 \sqrt{\pi}} 
			\operatorname{Erf}\left[\frac{\mu_l - \log\left[-1 + \epsilon + \epsilon \, \zeta	 \right]}{\sqrt{2}
			\sigma_l}\right],
			\label{EQ:C_S5}
		\end{align}
		%
		%
		where $\zeta = \exp{\left( \mu_e - \sqrt{2} \eta_n \sigma_e \right)}$. 
		Inserting \eqref{EQ:SECRECY_CAPACITY_EXISTENCE}, \eqref{EQ:C_S4} and \eqref{EQ:C_S5}   into \eqref{EQ:C_S3}, yields the secrecy outage probability as in \eqref{EQ:SECRECY_OUTAGE_PROBABILITY}.
\end{IEEEproof}
%

\section{Numerical Results}
%
%
In this section, we apply the framework to evaluate the feasibility of \ac{PHY} security in terms of the existence of secrecy capacity and the outage secrecy probability.
Eavesdropper and legitimate nodes are affected by shadowed fading with the \ac{LN} shadowing following a zero-mean Gaussian distribution with variance $\sigma=4$, Rician fading factor of $K=14.8\,\mathrm{dB}$, \review{and Hermite polynomial order of $N=24$ to evaluate our performance metrics.}
We consider that active nodes operate at a fixed transmit power of $20\,\mathrm{dBm}$.

Fig. \ref{FIG:PEX_VDIST_VMUX} depicts the
probability of existence of non-zero secrecy capacity  of the legitimate tagged link by varying its average \ac{SIR}, eavesdropper distance to the tagged transmitter and density
of interfering nodes for $d_l = 15 \mathrm{m}$.
We observe that the probability of non-zero
secrecy capacity increases when the average \ac{SIR} of the legitimate receiver
increases, whereas it decreases by positioning the eavesdropper close the
legitimate transmitter, namely, $d_e = 5 \mathrm{m}$ away.
In other words, the closer the eavesdropper is to the legitimate transmitter,
the higher should be the link quality of the legitimate pair so as to guarantee
its secrecy. 
\begin{figure}[!t]
	\centering
	\includegraphics[width=1\columnwidth]{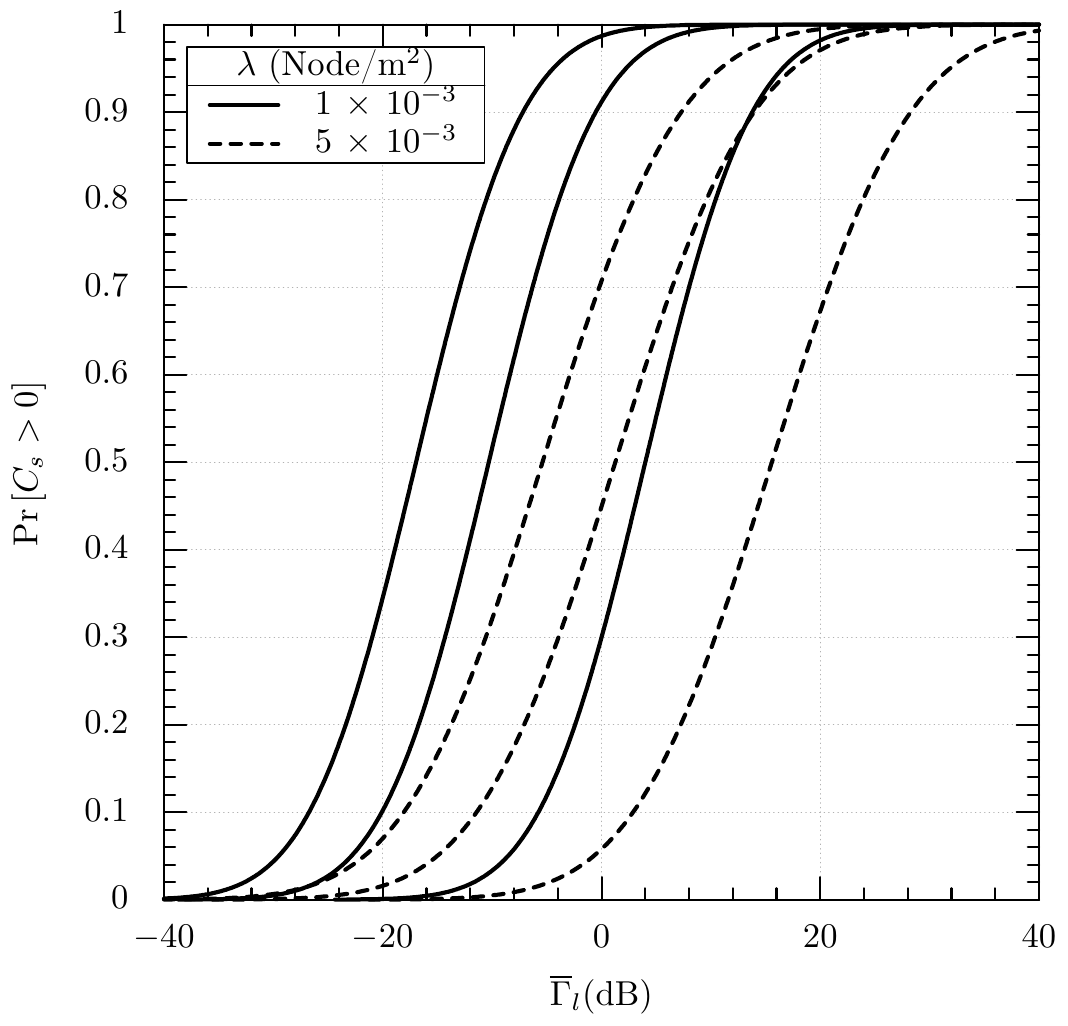}	
	\begin{tikzpicture}[overlay]
		\draw[o-o, black, line width = 1pt, dotted] (1.17, 6.5) to (2.55, 6.5)
		node[right = .1em, sloped] {$d_e = 5\,\mathrm{m}$};
		\draw[o-o, black, line width = 1pt, dotted] (-.4, 5.5) to (.9, 5.5)
		node[right = 1em, sloped] {$d_e = 15\,\mathrm{m}$};
		\draw[o-o, black, line width = 1pt, dotted] (-1.35, 4) to (-.2, 4)
		node[left = 3.0em, sloped] {$d_e = 25\,\mathrm{m}$};
	\end{tikzpicture}
	%
	%
	\caption{Probability of existence of non-zero secrecy capacity as a function of the
	perceived \ac{SIR} at the legitimate receiver for different distance
	between legitimate transmitter and eavesdropper  $d_e$, and density of interfering
	nodes $\lambda$, considering $d_l = 15 \, \mathrm{m}$.}
	\label{FIG:PEX_VDIST_VMUX}
	%
\end{figure}
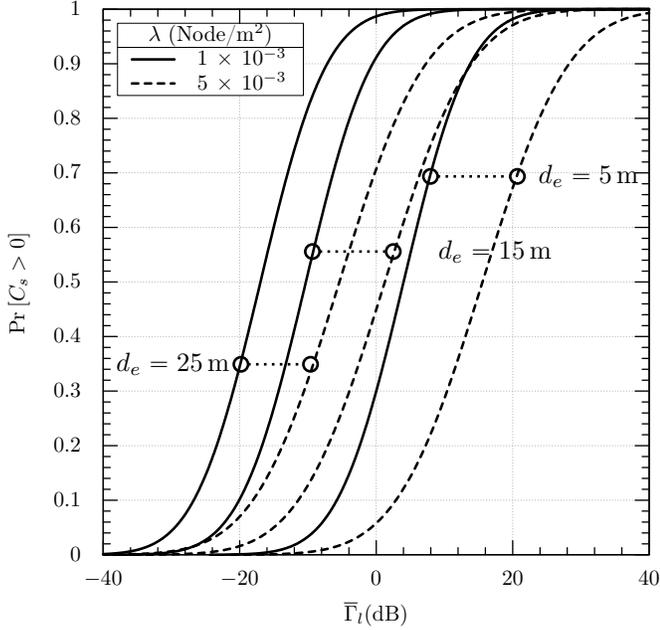
\begin{figure}[!t]
	\centering
	\includegraphics[width=1.\columnwidth]{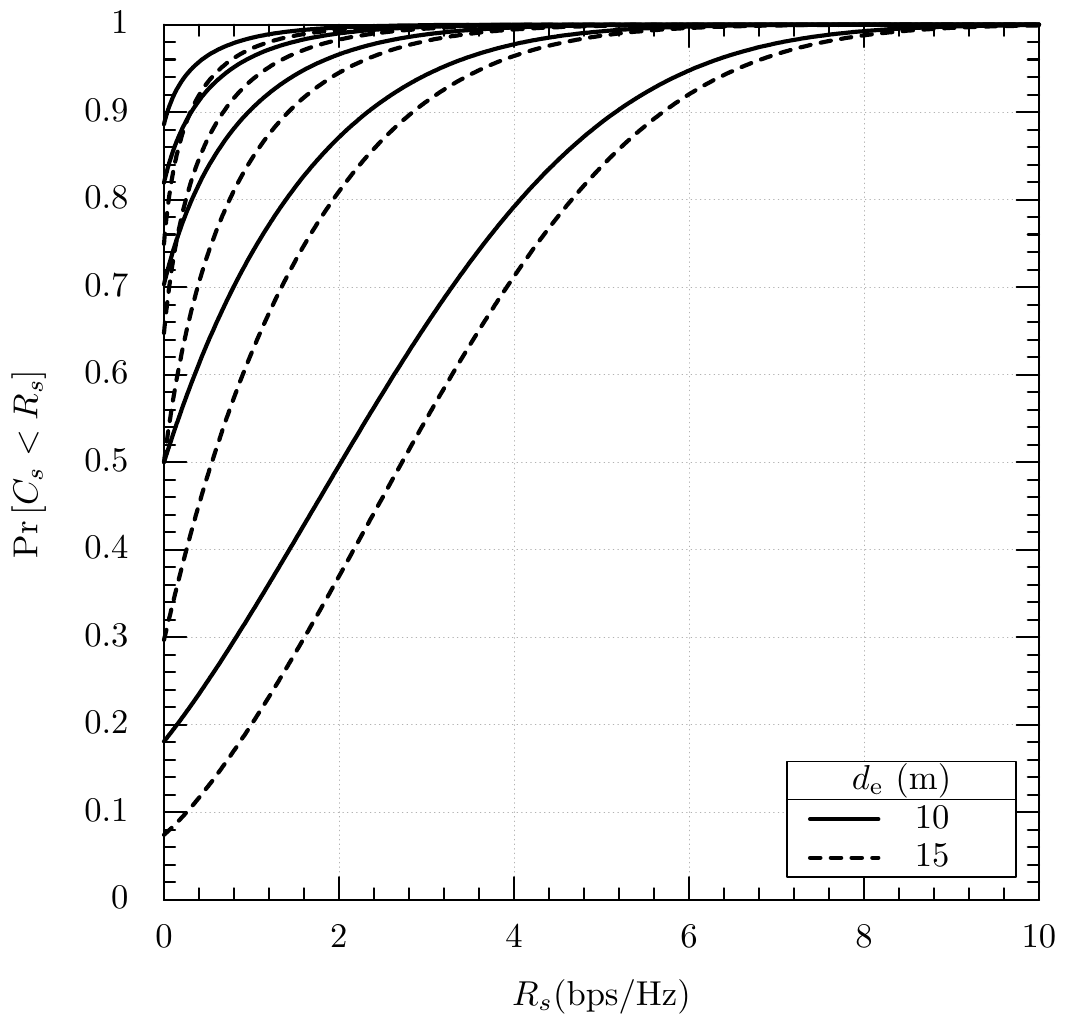}	
	\begin{tikzpicture}[overlay]
		\draw[-stealth', black, line width = 1pt] (0, 5.5) to (-3,8.5);
		\draw (1, 5) node[anchor = center] {$d_l = 5, 10, 15, 20,
		25\,\mathrm{m}$};
	\end{tikzpicture}
	%
	%
	\caption{Secrecy outage probability as a function of the secrecy rate for $d_l = 5, 10, 15, 20, 25\,\mathrm{m}$, while the eavesdropper is positioned $d_e=10\, \mathrm{m}$ and $d_e=15\,\mathrm{m}$ away from the transmitter.}
	\label{FIG:POUT_VRS}
	%
\end{figure}
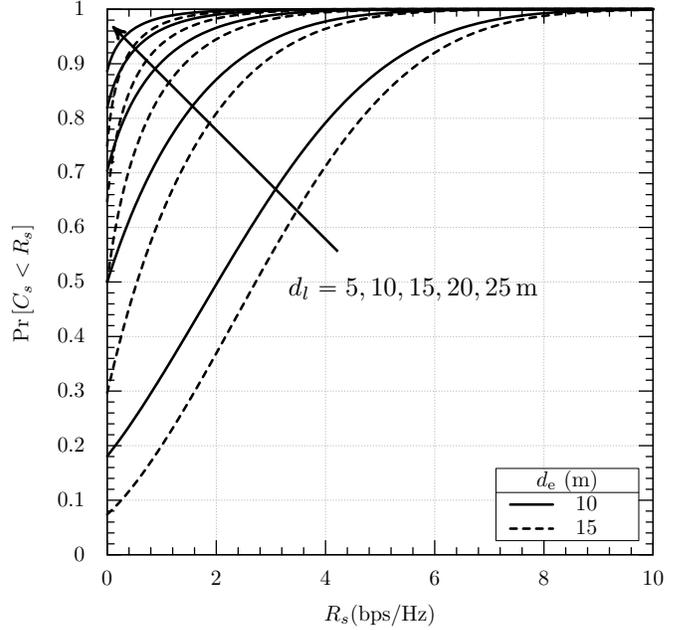

\review{
We can also see from Fig. \ref{FIG:PEX_VDIST_VMUX} that the increase of the interferers' density decreases the non-zero secrecy capacity probability for the same average \ac{SIR}, regardless of the distance $d_e$. This fact evinces that the increase of co-channel interference caused by increasing $\lambda$ equally decreases the channel capacity for both legitimate and eavesdropper links and therefore the gap between the curves with different densities tends to be constant when the same distance $d_e$ is set.}

Fig.~\ref{FIG:POUT_VRS} shows the secrecy outage probability for increasing
secrecy rate $R_s$, considering that the eavesdropper is at distance $d_e =10$ and  $d_e =15$ $\mathrm{m}$.
Note that the distance between the legitimate pair varies with increments of $5$~$\mathrm{m}$.
We can infer that secrecy outage probability worsens with the increase of the distance between the legitimate nodes and also with the proximity of the eavesdropper to the transmitter. 
However, higher secrecy rates can be achieved with shorter legitimate links once the transmitted signal is much stronger than the interference seen at the receivers. 

\section{Final remarks and conclusion} 

We investigated the secrecy capacity of the channel in large
scale, interference-limited networks. Using a more general
model that captures randomness due to interferers' position,
shadowing and fast fading, we derive closed form expressions
for the probability of non-zero secrecy capacity and secrecy
outage probability. For the proposed scenario, our numerical
results show under which conditions secrecy can be achieved
for different network configurations, evincing the effects of the
proximity of the eavesdroppers to the legitimate transmitter
and the interferer density on the secrecy outage probability.

We plan to continue the investigation introduced in this paper in scenarios legitimate transmitters use random access protocols and links are subject to quality constraints.
In this way, we plan to evaluate both the effective secrecy throughput and the network spatial secrecy throughput, as in \cite{NardelliTMC2014}.
\review{Another extension is to assess the secrecy capacity of Poisson distributed networks where multi-hop links are allowed \cite{NardelliTWC2012} such that the relative positions between relays (defined by some specific hopping strategy) and eavesdroppers are expect to impact the system performance.} 


\bibliographystyle{IEEEtran}

%
\end{document}

%% file: acronym.tex
\begin{acronym}[]
	\acro{3GPP}[$3$GPP]{$3^\textnormal{rd}$ Generation Partnership Project}
	\acro{ABS}{Almost Blank Sub-frame}
	\acro{RAT}{Radio Access Technology}
	\acro{ADSL}{Asymmetric Digital Subscriber Line}
	\acro{CRS}{Cell-specific Reference Signal}
	\acro{KPI}{Key Performance Indicators}
	\acro{BC}{Broadcast Channel}
	\acro{ES}{Evaluation Scenario}
	\acro{RSS}{Received Signal Strength}
	\acro{RNTP}{Relative Narrowband Transmit Power}
	\acro{RSRP}{Reference Signal Received Power}
	\acro{OS}{Order Statistic}
	\acro{SF}{Sub-Frame}
	\acro{ALBA-R}{Adaptive Load-Balanced Algorithm Rainbow}
	\acro{ALBA}{Adaptive Load-Balanced Algorithm}
	\acro{ALOHA}[ALOHA]{}
	\acro{APDL}{Average Packet Delivery Latency}
	\acro{AP}{Access Point}
	\acro{ASE}{Area Spectral Efficiency}
	\acro{BAP}{Blocked Access Protocol}
	\acro{BB}{Busy Burst}
	\acro{BPP}{Binomial Point Process}
	\acro{BS}{Base Station}
	\acro{CAA}{Channel Access Algorithm}
	\acro{CAP}{Channel Access Protocol}
	\acro{CCDF}{Complementary Cumulative Distribution Function}
	\acro{CCI}{Co-Channel Interference}
	\acro{CDF}{Cumulative Distribution Function}
	\acro{CDMA}{Code Division Multiple Access}
	\acro{CDR}{Convex Lenses Decision Region}
	\acro{CF}{Characteristic Function}
	\acro{CGF}{Contention-based Geographic Forwarding}
	\acro{CM}{Coordination Mechanism}
	\acro{CPICH}{Common Pilot Channel}
	\acro{CRA}{Conflict Resolution Algorithm}
	\acro{CRD}{Contention Resolution Delay}
	\acro{CRI}{Contention Resolution Interval}
	\acro{CRP}{Contention Resolution Protocol}
	\acro{CR}{Contention Resolution}
	\acro{CSMA/CA}{Carrier Sense Multiple Access with Collision Avoidance}
	\acro{CSMA/CD}{Carrier Sense Multiple Access with Collision Detection}
	\acro{CSMA}{Carrier Sense Multiple Access}
	\acro{CTM}{Capetanakis-Tsybakov-Mikhailov}
	\acro{CTS}{Clear To Send}
	\acro{DAS}{Distributed Antenna System}
	\acro{DCF}{Distributed Coordination Function}
	\acro{DER}{Dynamic Exclusion Region}
	\acro{DL}{Downlink}
	\acro{E2E}{End-to-End}
	\acro{EDM}{Euclidean Distance Matrix}
	\acro{FAP}{Femto Access Point}
	\acro{FBS}{Femto Base Station}
	\acro{FDD}{Frequency Division Duplexing}
	\acro{FDM}{Frequency Division Multiplexing}
	\acro{FDR}{Forwarding Decision Region}
	\acro{FFR}{Fractional Frequency Reuse}
	\acro{FG}{Frequency Group}
	\acro{FPP}{First Passage Percolation}
	\acro{FUE}{Femto User Equipment}
	\acro{FU}{Femtocell User}
	\acro{GF}{Geographic Forwarding}
	\acro{GLIDER}{Gradient Landmark-Based Distributed Routing}
	\acro{GPSR}{Greedy Perimeter Stateless Routing}
	\acro{GeRaF}{Geographic Random Forwarding}
	\acro{HDR}[HDR]{High Data Rate}
	\acro{HII}{High Interference Indicator}
	\acro{HNB}{Home Node B}
	\acro{HOS}{Higher Order Statistics}
	\acro{HUE}{Home User Equipment}
	\acro{ICIC}{Inter-Cell Interference Coordination}
	\acro{IEEE}[IEEE]{}
	\acro{IP}{Interference Profile}
	\acro{LN}{Log-Normal}
	\acro{LTE}{Long Term Evolution}
	\acro{LoS}{Line-of-Sight}
	\acro{MACA}{Multiple Access with Collision Avoidance}
	\acro{MAC}{Medium Access Control}
	\acro{MBS}{Macro Base Station}
	\acro{MGF}{Moment Generating Function}
	\acro{MIMO}{Multiple-Input Multiple-Output}
	\acro{MPP}{Marked Point Process}
	\acro{MRC}{Maximum Ratio Combining}
	\acro{MS}{Mobile Station}
	\acro{MUE}{Macro User Equipment}
	\acro{MU}{Macrocell User}
	\acro{NB}{Node B}
	\acro{NLoS}{Non Line-of-Sight}
	\acro{NRT}{Non Real Time}
	\acro{OFDMA}{Orthogonal Frequency Division Multiple Access}
	\acro{OOP}{Object Oriented Programming}
	\acro{OP}{Outage Probability}
	\acro{PBS}{Pico Base Station}
	\acro{PC}{Power Control}
	\acro{PDF}{Probability Distribution Function}
	\acro{PDSR}{Packet Delivery Success Ratio}
	\acro{PGF}{Probability Generating Function}
	\acro{PHY}{Physical Layer}
	\acro{PMF}{Probability Mass Function}
	\acro{PPP}{Point Process}
	\acro{PPP}{Poisson Point Process}
	\acro{PRM}{Poisson Random Measure}
	\acro{QoS}{Quality of Service}
	\acro{RAS}{Random Access System}
	\acro{RA}{Random Access}
	\acro{RCA}{Random Channel Access}
	\acro{REB}{Range Expansion Bias}
	\acro{RE}{Range Expansion}
	\acro{RF}{Radio Frequency}
	\acro{RIBF}{Regularized Incomplete Beta Function}
	\acro{RMA}{Random Multiple-Access}
	\acro{RRM}{Radio Resource Management}
	\acro{RSA}{Relay Selection Algorithm}
	\acro{RS}{Relay Selection}
	\acro{RTS}{Request to Send}
	\acro{RT}{Real Time}
	\acro{RV}{Random Variable}
	\acro{SC}{Selection Combining}
	\acro{SDR}{Sectoral Decision Region}
	\acro{SG}{Stochastic Geometry}
	\acro{SIC}{Successive Interference Cancellation}
	\acro{SINR}{Signal-to-Interference plus Noise Ratio}
	\acro{SIR}{Signal-to-Interference Ratio}
	\acro{SLN}{Shifted Log-Normal}
	\acro{SNR}{Signal to Noise Ratio}
	\acro{SON}{Self-Organizing Network}
	\acro{SPP}{Spatial Poisson Process}
	\acro{STA}{Standard Tree Algorithm}
	\acro{TAS}{Transmit Antenna Selection}
	\acro{TC}{Transmission Capacity}
	\acro{TDD}{Time Division Duplexing}	
	\acro{TDMA}{Time Division Multiple Access}	
	\acro{TS}{Terminal Station}
	\acro{UDM}{Unit Disk Model}
	\acro{UD}{Unit Disk}
	\acro{UE}{User Equipment}
	\acro{ULUTRANSIM}[UL UTRANSim]{R6 Uplink UTRAN Simulator}
	\acro{UL}{Uplink}
	\acro{UML}{Unified Modeling Language}
	\acro{UMTS}{Universal Mobile Telecommunications System}
	\acro{WCDMA}{Wideband Code Division Multiple Access}
	\acro{WSN}{Wireless Sensor Network}
	\acro{iid}[\textup{i.i.d.}]{independent and identically distributed}
\end{acronym}